\documentclass[10pt,conference]{IEEEtran}
\IEEEoverridecommandlockouts
\usepackage{cite}
\usepackage{amsmath,amssymb,amsfonts}
\usepackage{algorithmic}
\usepackage{graphicx}
\usepackage{textcomp}
\usepackage{hyperref}
\usepackage{xcolor}
\usepackage[utf8]{inputenc}
\usepackage{mathrsfs}  
\usepackage{nicefrac}       
\usepackage{graphicx}
\usepackage{booktabs}
\usepackage{amsthm}
\usepackage{mathtools}
\usepackage[tableposition=top]{caption}
\usepackage{xcolor}
\usepackage{wrapfig}
\usepackage{siunitx}
\usepackage{subfigure}
\theoremstyle{remark}
\newtheorem*{remark}{Remark}
\DeclareMathOperator{\Tr}{Tr}
\DeclareMathAlphabet\mathbfcal{OMS}{cmsy}{b}{n}
\def\BibTeX{{\rm B\kern-.05em{\sc i\kern-.025em b}\kern-.08em
    T\kern-.1667em\lower.7ex\hbox{E}\kern-.125emX}}
\begin{document}

\title{ML-aided power allocation for Tactical MIMO \\
\thanks{Research was sponsored by the Army Research Office and was accomplished under Cooperative Agreement Number W911NF-19-2-0269. 
The views and conclusions contained in this document are those of the authors and should not be interpreted as representing the official policies, either expressed or implied, of the Army Research Office or the U.S. Government. 
The U.S. Government is authorized to reproduce and distribute reprints for Government purposes notwithstanding any copyright notation herein.}
}

\author{\IEEEauthorblockN{Arindam Chowdhury\IEEEauthorrefmark{1}, Gunjan Verma\IEEEauthorrefmark{2}, Chirag Rao\IEEEauthorrefmark{2}, Ananthram Swami\IEEEauthorrefmark{2} and Santiago Segarra\IEEEauthorrefmark{1}} 
\IEEEauthorblockA{\IEEEauthorrefmark{1}Electrical \& Computer Engineering, Rice University, USA\\
Email: \{ac131, segarra\}@rice.edu} \IEEEauthorblockA{\IEEEauthorrefmark{2}US Army DECVOM Army Research Laboratory, USA\\
Email: \{gunjan.verma.civ, chirag.r.rao.civ, ananthram.swami.civ\}@army.mil}}

\maketitle

\begin{abstract}
We study the problem of optimal power allocation in single-hop multi-antenna ad-hoc wireless networks. 
A standard technique to solve this problem involves optimizing a tri-convex function under power constraints using a block-coordinate-descent based iterative algorithm. This approach, termed WMMSE, tends to be computationally complex and time consuming. 
Several learning-based approaches have been proposed to speed up the power allocation process. 
A recent work, UWMMSE, learns an affine transformation of a WMMSE parameter in an unfolded structure to accelerate convergence. 
In spite of achieving promising results, its application is limited to single-antenna wireless networks. 
In this work, we present a UWMMSE framework for power allocation in (multiple-input multiple-output) MIMO interference networks.
A major advantage of this method lies in its use of low-complexity learnable systems in which the number of parameters scales linearly with respect to the hidden layer size of embedded neural architectures and the product of the number of transmitter and receiver antennas only, fully independent of the number of transceivers in the network.   
We illustrate the superiority of our method through an empirical study of our approach in comparison to WMMSE and also analyze its robustness to changes in channel conditions and network size.          
\end{abstract}


\section{Introduction}
The future tactical wireless communications network must cope with an increasingly congested and contested spectrum environment, and under adversarial conditions and high tempo as envisaged in multi-domain operations (MDO)~\cite{army2018us}. 
The emerging area of massive MIMO is a potential solution to increase rate and survivability~\cite{bjornson2019massive,kekirigoda2019massive}, with increasing DoD interest in adapting commercial 5G solutions to tactical network challenges \cite{Harvey2019mmwave}, \cite{defense.gov20215Gtestbeds}. 
Concomitant with this are challenges related to computational complexity of implementing (near-)optimal solutions, as the number of devices and/or number of antennas increase. 
Domain-aware machine learning algorithms have recently emerged as promising solution approaches~\cite{hu2020iterative,pellaco2020deep,chowdhury2021unfolding,kumar2021adaptive,zhao2021distributed}. 
The idea here is to combine relevant domain information available in the form of a structured algorithmic solution for a specific problem with neural architectures that learn directly from data~\cite{monga2019algorithm,shlezinger2020model}. 
These data-driven hybrid connectionist models leverage the universal approximation property of neural networks to learn a suitable functional transformation under the framework of a classical method to achieve a better convergence. 
Some of these techniques have been successfully applied in designing model-aware deep receivers to carry out symbol detection and recovery in (single-input single-output) SISO as well as MIMO systems~\cite{samuel2019learning,shlezinger2020viterbinet,farsad2020data,pratik2020re}. 
Another crucial problem that has been investigated within this framework is that of power allocation in wireless networks through interference management~\cite{sun2018learning,lee2018deep,eisen2019learning,shlezinger2020deepsic}.   

Optimal power allocation in wireless networks, especially under fast-changing channel conditions, is a challenging task that requires solving a non-convex constrained sum-rate optimization problem. 
The classical solution approach is to reformulate it as a weighted minimum mean square error~\cite{shi2011iteratively} (WMMSE) optimization problem. 
The WMMSE objective is tri-convex, i.e., convex in each of its three variables when the others are fixed and, therefore, more tractable than the original problem. 
However, its performance is limited by the cumbersome iterative steps which tend to be computationally expensive (quadratic in the number of users) and time consuming. This is especially detrimental for fast-changing channels. 
UWMMSE-SISO~\cite{chowdhury2021unfolding} addressed this drawback for a single-antenna ad-hoc interference network by truncating WMMSE iterations and embedding a graph neural network (GNN) based learnable component within the structure to compensate for the truncation. 
However, this method in its original form cannot handle MIMO systems. On the other hand, WMMSE-PGD~\cite{pellaco2020deep} and IAIDNN~\cite{hu2020iterative} solve the same problem using distinct unfolding schemes that seek to reduce computational complexity by replacing expensive matrix operations using learnable components for MISO and MIMO systems respectively. 
Nevertheless, neither WMMSE-PGD nor IAIDNN preserve the original block-coordinate-descent (BCD) update structure that UWMMSE-SISO captures. 
Further, the use of GNNs allows it to leverage the underlying topology of wireless network graphs and confers the architecture with permutation equivariance, thereby enhancing generalizability. 
In this work, we design an unfolding scheme for WMMSE to make it amenable to MIMO systems. 

\vspace{1mm}\noindent\textbf{Contribution.} 
The contributions of this paper are twofold:\\
i) We propose UWMMSE-MIMO architecture for power allocation in multi-antenna interference scenarios.\\
ii) We empirically evaluate the performance of the proposed method in comparison to WMMSE, demonstrate its generalization to networks of unseen sizes, and illustrate its robustness with respect to varying channel distributions.

\vspace{1mm}\noindent\textbf{Notation.}
The entries of a multi-dimensional tensor $\mathbfcal{X}$, a matrix $\mathbf{X}$, and a vector $\mathbf{x}$ are denoted by $[\mathbfcal{X}]_{ij\dots}$, $[\mathbf{X}]_{ij}$, and $[\mathbf{x}]_{i}$. 
The generic subindex $:$ denotes a whole dimension, e.g., $[\mathbf{X}]_{i \,:}$ refers to row $i$ of matrix $\mathbf{X}$.

\section{System Model and Problem Formulation}
Consider a single-hop ad-hoc interference network with $M$ distinct transmitter-receiver pairs. 
Each transmitter is equipped with $T$ antennas while the number of receiver antennas is $R$. 
Let $\mathbf{V}_{i} \in \mathbb{R}^{T\times d}$ denote the beamformer that transmitter $i$ uses to transmit a signal $\mathbf{s}_i \in \mathbb{R}^{d}$  to its corresponding receiver $r(i)$. 
Assuming a linear channel model, the received signal $\mathbf{y}_i \in \mathbb{R}^{R}$, at $r(i)$ is given by
\begin{equation}\label{E:trans_model}
    \mathbf{y}_i = \underbrace{\mathbf{H}_{ii}\mathbf{V}_i\mathbf{s}_i}_{desired\ signal} + \underbrace{\sum_{\substack{j=1 \, | \,  j\neq i}}^M \mathbf{H}_{ij}\mathbf{V}_j\mathbf{s}_j + \mathbf{n}_i}_{interference\ plus\ noise}, \quad \forall i,
\end{equation}
where $\mathbf{H}_{ij} \in \mathbb{R}^{R\times T}$ represents the channel from transmitter $i$ to receiver $r(j)$, while $\mathbf{n}_i \in \mathbb{R}^{R}$ denotes additive white Gaussian noise sampled from a standard normal distribution $\mathscr{N}(0,\sigma^2\mathbf{I})$. 
Finally, the transmitted signal is estimated at $r(i)$ using a receiver beamformer $\mathbf{U}_{i} \in \mathbb{R}^{R\times d}$, to obtain $\Hat{\mathbf{s}}_i = \mathbf{U}_i^\top \mathbf{y}_i$ for all $i \in {\{1, \dots, M\}}$.

The performance of a wireless network, such as the one described above, is measured in terms of certain utilities of the system. 
For instance, the capacity of the network is defined as a weighted combination of the instantaneous data rate $c_i$ achievable at each receiver $r(i)$, given by Shannon's theorem as a function of the signal-to-interference-plus-noise-ratio.
More precisely, if we define the channel state information (CSI) tensor $\mathbfcal{H} \in \mathbb{R}^{M\times M\times R\times T}$ such that $[\mathbfcal{H}]_{ij::} = \mathbf{H}_{ij}$ and the transmitter beamformer tensor $\mathbfcal{V} \in \mathbb{R}^{M\times T\times d}$ such that $[\mathbfcal{V}]_{i::} = \mathbf{V}_{i}$, we have that for every user $i$ its rate is given by
\begin{align}\label{E:data_rate}
    c_i(\mathbfcal{V},\mathbfcal{H}) &\coloneqq \log_2 \det \big( ~~ \mathbf{I} + \mathbf{H}_{ii}^{}\mathbf{V}_{i}^{} \mathbf{V}_{i}^{\top} \mathbf{H}_{ii}^{\top} \nonumber \\ 
    & \quad \quad \quad \quad  \left( \sigma^2\mathbf{I} + \sum_{j\neq i} \mathbf{H}_{ij}^{}\mathbf{V}_{j}^{} \mathbf{V}_{j}^{\top} \mathbf{H}_{ij}^{\top} \right)^{-1} \big) .
\end{align}
Hence, our objective is to determine the transmit beamformer $\mathbfcal{V}$ that maximizes the network capacity under a given power constraint
\begin{align}\label{E:optimization_problem}
		 &\quad \quad \max_{\mathbfcal{V}} \,\, \sum_{i=1}^M \alpha_i c_i(\mathbfcal{V}, \mathbfcal{H}) \quad  \\& \text{s.t.} \,\,\,\,\,  \Tr\left(\mathbf{V}_i^{}\mathbf{V}_i^{\top}\right) \leq P_{\mathrm{max}}, \,\,\,\, \forall \,\, i. \nonumber
\end{align}
Henceforth, for simplicity we focus on the case where every user is given the same weight $\alpha_i=1$ in the objective. However, the optimization problem in~(\ref{E:optimization_problem}) is non-convex and has been shown to be NP-hard~\cite{luo_2008_dynamic,hong_2014_signal}. 
The standard approach to solve this problem is to reformulate it as a constrained WMMSE optimization problem~\cite{shi2011iteratively}.
Specifically, introducing the tensors $\mathbfcal{W} \in \mathbb{R}^{M\times d\times d}$ and $\mathbfcal{U} \in \mathbb{R}^{M\times R\times d}$ we have that
\begin{align}\label{E:problem_reformulation}
&\min_{\mathbfcal{W,U,V}} \sum_{i=1}^M (\Tr(\mathbf{W}_i \mathbf{E}_i) - \log \det \mathbf{W}_i ),\\
& \text{s.t.} \,\,\,\,\,  \Tr\left(\mathbf{V}_i^{}\mathbf{V}_i^{\top}\right) \leq P_{\mathrm{max}}, \,\,\,\, \forall \,\, i, \nonumber
\end{align}
where $\mathbf{W}_i = [\mathbfcal{W}]_{i::} \succeq 0$ is a weight matrix at receiver $r(i)$ while $\mathbf{E}_i$ is the mean square error between transmitted and received signals, which depends on $\mathbfcal{V}$ and $\mathbfcal{U}$. 

The optimization problem~\eqref{E:problem_reformulation} is equivalent to~\eqref{E:optimization_problem} as it can be shown~\cite[Thm. 3]{shi2011iteratively} that the variable $\mathbfcal{V}^*$ in the global optimal solution $\{\mathbfcal{W}^*, \mathbfcal{U}^*, \mathbfcal{V}^*\}$ of the former is the same as the optimal transmit beamformer $\mathbfcal{V}^*$ in the latter. 
Moreover, the problem in~\eqref{E:problem_reformulation} is \emph{tri-convex}, i.e., the objective is convex in each of the three variables when the other two are fixed.
This makes~\eqref{E:problem_reformulation} amenable to a BCD solution, which we leverage in Section~\ref{sec:uwmmse_mimo} for our unfolded architecture.
%
%
In spite of this tri-convexity, WMMSE performance is limited by its cumbersome iterative steps which are computationally complex (per-iteration complexity scales $\mathcal{O}(M^2)$ where $M$ is the number of users)~\cite{chowdhury2021unfolding}. 
It can also be time consuming depending on the size and complexity of the wireless network, making it particularly ineffective for fast-changing channels.

\section{Unfolded WMMSE for MIMO}
\label{sec:uwmmse_mimo}

Algorithm unfolding~\cite{liu2019deep,monga2019algorithm,farsad2020data} refers to constructing a learnable connectionist architecture using the update structure of an iterative algorithm. 
The process essentially involves truncating the iterations of a classical algorithm to form a cascade of a limited number of \emph{layers} each of which copies the update operations of the algorithm in addition to carrying a neural module that approximates one or more parameters of the algorithm. 
In this way, an unfolded algorithm synergistically combines the domain information available in the form of a structured solution method for a given problem with neural-network-based modules that can be learned from data.   

\begin{figure}[t]
	\centering
	\includegraphics[width=\linewidth]{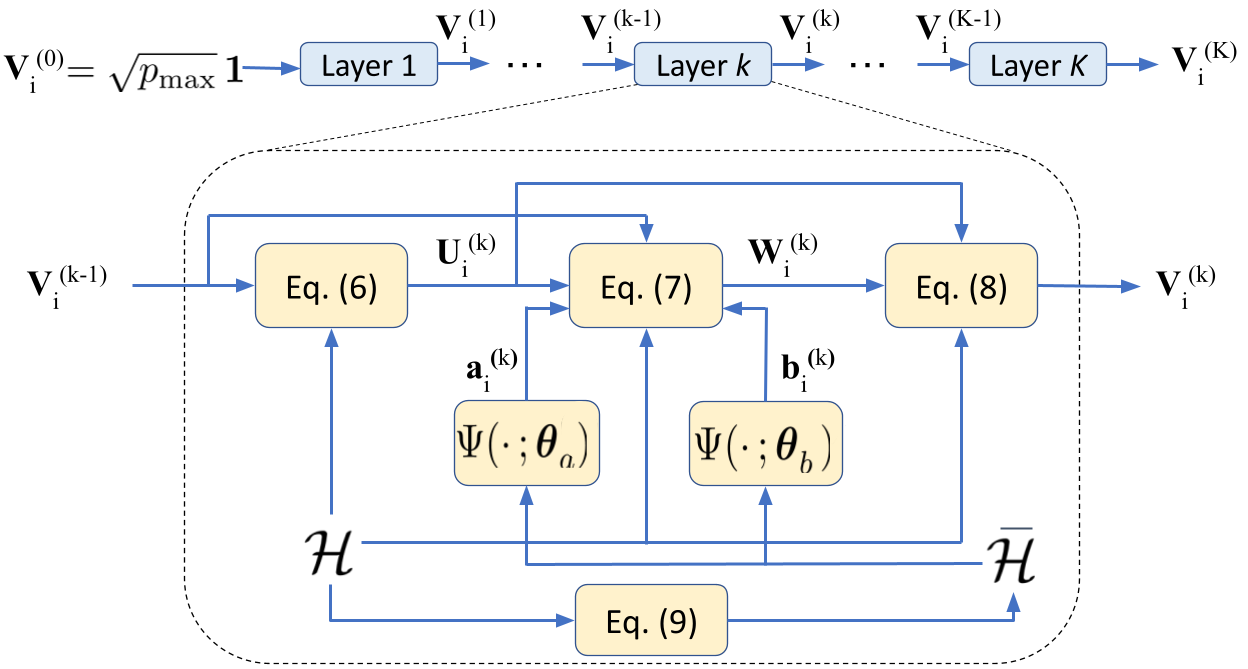}
	\vspace{-0.4cm}
	\caption{\small Schematic diagram of the proposed UWMMSE algorithm for MIMO wireless networks.
	A generic intermediate layer $k$ is detailed.}
	\label{F:unr}
\end{figure}

Recalling that our goal is to solve~\eqref{E:optimization_problem}, we seek a parametric function $\Lambda( \cdot ; \Theta): \mathbb{R}^{M\times M\times R\times T} \to \mathbb{R}^{M\times T\times d}$ where $\Theta$ contains learnable parameters and $\mathbfcal{V} = \Lambda( \mathbfcal{H} ; \Theta)$ approximates the solution to~\eqref{E:optimization_problem} for given CSI $\mathbfcal{H}$.
In this work, we propose a neural network architecture for $\Lambda$ inspired by the unfolding of WMMSE into a set of hybrid layers each of which carries an embedded graph neural network (GNN) learning module to accelerate convergence. 
More specifically, by setting the matrices $\mathbf{V}^{(0)}_i = \sqrt{P_\mathrm{max}} \,  \mathbf{1}_{T \times d}$ for every user $i$, we have that for layers $k = 1, ... K$,
\begin{equation}
 \hspace{4mm} \mathbf{a}^{(k)} = \Psi({\mathbfcal{\bar{H}}}; \theta_a^{}), \quad \mathbf{b}^{(k)} = \Psi({\mathbfcal{\bar{H}}}; \theta_b^{}), \qquad  \hspace{-5mm} \label{E:unfold_3} 
\end{equation}
\begin{equation}
 \mathbf{U}^{(k)}_i \!\!=\!\! \bigg(\!\sum_{\substack{j\neq i}}\mathbf{H}_{ij}^{}\mathbf{V}_{j}^{(k-1)}\mathbf{V}_{j}^{(k-1)\top}\!\mathbf{H}_{ij}^{\top} \!+ \!\sigma^2\mathbf{I}\!\bigg)^{-1}\!\!\!\!\mathbf{H}_{ii}^{}\mathbf{V}_{i}^{(k-1)}\!\! \label{E:unfold_u} 
\end{equation}
\begin{equation}
\mathbf{W}^{(k)}_i \!=\! [\mathbf{a}^{(k)}]_i \big( \mathbf{I} - \mathbf{U}_i^{(k)\top}\mathbf{H}_{ii}^{}\mathbf{V}_{j}^{(k-1)} \big)^{-1} + [\mathbf{b}^{(k)}]_i \label{E:unfold_w}
\end{equation}
\begin{equation}
\mathbf{V}^{(k)}_i \! = \! \beta \Bigg( \!\! \bigg( \sum_{\substack{j\neq i}} \mathbf{H}_{ij}^{\top}\mathbf{U}_j^{(k)}\mathbf{W}_j^{(k)}\mathbf{U}_j^{(k)\top}\mathbf{H}_{ij}^{} \bigg)^{-1}\!\!\! \mathbf{H}_{ii}^{\top}\mathbf{U}_i^{(k)}\mathbf{W}_i^{(k)} \! \!\Bigg)\!\label{E:unfold_v}
\end{equation}
%
where updates \eqref{E:unfold_u}-\eqref{E:unfold_v} are repeated for every user $i$, $\Psi$ in~\eqref{E:unfold_3} is a GNN architecture with a set of trainable weights $\theta$, {$\mathbfcal{\bar{H}}$} is a transformation to the CSI to be described in~\eqref{eq:H_bar}, and $\beta (\cdot)$ in~\eqref{E:unfold_v} is a non-linear activation to be detailed in~\eqref{eq:nonlinear_beta}.
The output of our layered architecture is then given by $\mathbfcal{V} = \Lambda( \mathbfcal{H} ; \Theta)$ such that $[\mathbfcal{V}]_{i::} = \mathbf{V}^{(K)}_i$ for all $i$.
A schematic view of the variable dependence of UWMMSE is given in Fig.~\ref{F:unr}.

The first notable point about the UWMMSE equations in \eqref{E:unfold_3}-\eqref{E:unfold_v} is that if we ignore~\eqref{E:unfold_3} and instead set $\mathbf{a}^{(k)} = \mathbf{1}$ and $\mathbf{b}^{(k)} = \mathbf{0}$ for every layer $k$, then \eqref{E:unfold_u}-\eqref{E:unfold_v} boil down to the classical BCD closed-form updates of WMMSE.
However, by providing the additional flexibility to UWMMSE of learning $\mathbf{a}^{(k)} \in \mathbb{R}^{M}$ and $\mathbf{b}^{(k)} \in \mathbb{R}^{M}$ in \eqref{E:unfold_3} -- which are implemented as weights of an affine transform in~\eqref{E:unfold_w} --, we enable faster convergence and better performance compared with the classical WMMSE, as we illustrate in Section~\ref{sec:num_exp}.

%

\begin{figure}[t]
	\centering
	\includegraphics[width=\linewidth,height=0.50\linewidth]{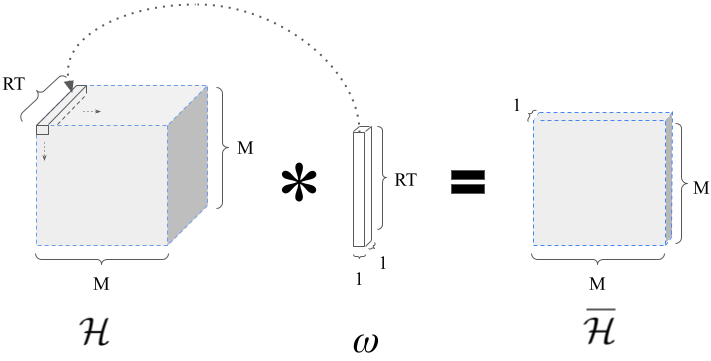}
	\vspace{-0.4cm}
	\caption{\small Operational structure of $\Phi$ in~\eqref{eq:H_bar}. A shared $1\times1$ filter kernel $\omega$ generates scalar representations for all $M^2$ elements of $\mathbfcal{H}$.} 
	\label{F:phi}
\end{figure}

The choice of the specific GNN $\Psi$ in~\eqref{E:unfold_3} and the size of the trainable weights $\theta$ are arbitrary and can be made depending on the nature of the problem. 
Independent of the choice of architecture, we treat the channel matrix as a weighted adjacency matrix of a directed graph which is used to aggregate information from the neighboring nodes~\cite{segarra_2017_optimal}. 
In the multi-antenna setting that we consider here, the channel between a transmitter $i$ and a receiver $r(j)$ is described by an $R \times T$ matrix that depends on the the number of transmitter and receiver antennas. 
However, since GNN $\Psi$ requires the CSI between $i$ and $r(j)$ to be represented by a scalar~\cite{kipf2016semi}, we propose the use of a single-layered $1\times 1$ depth-wise convolution operation with shared filter weights to reduce $\mathbfcal{H}$ to an amenable structure.
Essentially, we define an additional fully connected neural layer $\Phi(\mathbfcal{H};\omega): \mathbb{R}^{M\times M\times R \times T} \to \mathbb{R}^{M\times M\times 1}$ where $\omega \in \mathbb{R}^{RT}$ and 
\begin{equation}\label{eq:H_bar}
{\mathbfcal{\bar{H}}} = \Phi(\mathbfcal{H};\omega)  
\end{equation}
forms the input to $\Psi$ in~\eqref{E:unfold_3}\footnote{In practice, we observed that a row-normalization operation on $\mathbfcal{\bar{H}}$ prior to inputting it in~\eqref{E:unfold_3} had the effect of stabilizing training and thereby improving the overall performance. See footnote 2 for access to implementation code.}. Figure~\ref{F:phi} illustrates the operational structure of $\Phi$ where we first reshape the last two dimensions in $\mathbfcal{H}$ into a single dimensions of size $RT$ and then apply a convolution.
Indeed, this operation can be interpreted as a learnable weighted combination of the $RT$ antenna coefficients for each channel matrix $\mathbf{H}_{ij} = [\mathbfcal{H}]_{ij}$ to generate a scalar representation $H_{ij}$ for the channel. 
It is important to note that the combining set of weights $\omega$ are identical for all channel elements and learned during training. 
This is appropriate as all channel representations must have identical functional mapping from their respective antenna coefficients in a way that is analogous to shared $1\times 1$ convolutions of image pixels. 
Additionally, having a shared filter kernel allows for an $\mathcal{O}(M^2)$ reduction in the number of learnable weights. 

\begin{figure*}[!htbp]
	\centering
	\subfigure[]{
			\centering
			\includegraphics[width=0.31\textwidth]{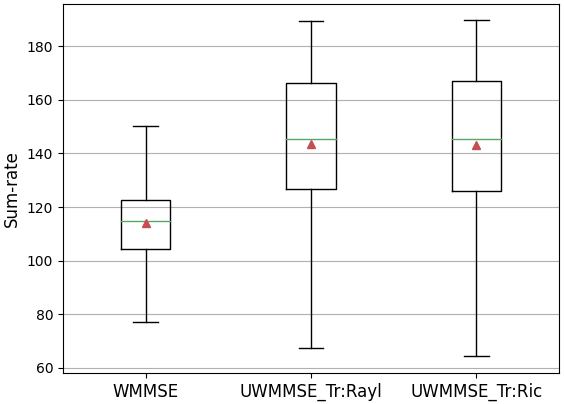}
			\label{Fig:ro1}
		}	
	\subfigure[]{
			\centering
			\includegraphics[width=0.31\textwidth]{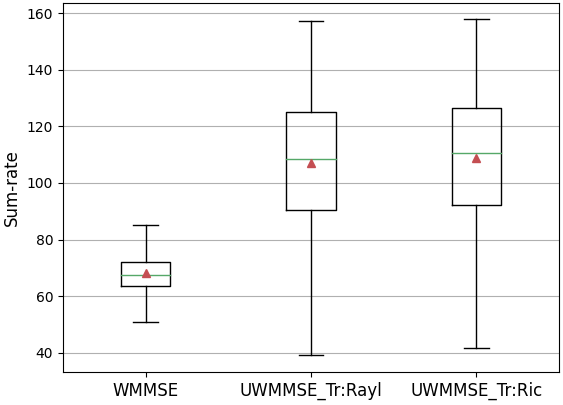}
			\label{Fig:ro2}
		}	
	\subfigure[]{
			\centering
			\includegraphics[width=0.31\textwidth,height=0.22\textwidth]{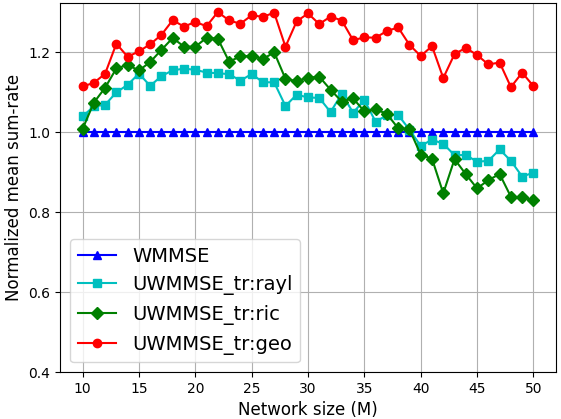}
			\label{Fig:ro3}
		}
		\caption{\small {Robustness analysis for UWMMSE. 
		(a)~Box plots of the achieved utilities for $10000$ randomly drawn \emph{Rayleigh} channel state matrices for $M = 20$ users. UWMMSE trained on \emph{Rayleigh} compared with UWMMSE trained on \emph{Rician} along with WMMSE.
		(b)~Box plots of the achieved utilities for $10000$ randomly drawn \emph{Rician} channel state matrices for $M = 20$ users. UWMMSE trained on \emph{Rician} compared with UWMMSE trained on \emph{Rayleigh} along with WMMSE.
		(c)~Mean sum-rate achieved on a \emph{Geometric} channel as network size $M$ varies in $\{10, \ldots, 50\}$. UWMMSE trained on \emph{Geometric} compared with UWMMSE trained on \emph{Rayleigh} and \emph{Rician} along with WMMSE.
		}}
		\label{Fig:impact}
\end{figure*}

The proposed architecture has very few trainable weights making it easy to train, and likely to generalize as illustrated by the numerical examples (see Fig~\ref{Fig:impact}). 
The number of weights $\theta$ in each of the two $2$-layered GCNs~\cite{kipf2016semi}, $\Psi$ is $6h + 2$, where $h$ is the size of the hidden layer; 
Further, the linear layer $\Phi$ has $RT + 1$ parameters, where $R$ and $T$ represent number of receive and transmit antennas. The total number of trainable weights is thus $12h + RT + 5$, and is independent of the number of users $M$.  

%

\begin{figure*}[!htbp]
	\centering
	\subfigure[]{
			\centering
			\includegraphics[width=0.31\textwidth]{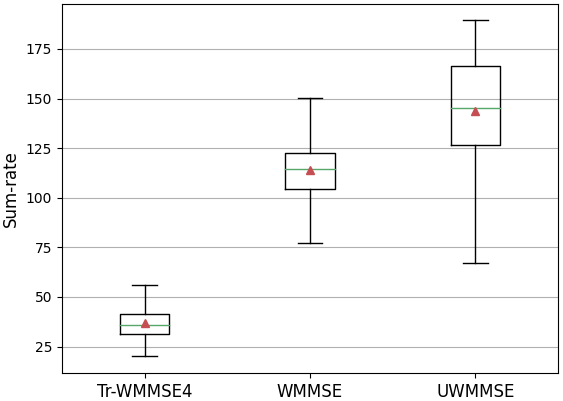}
			\label{Fig:ray}
		}	
	\subfigure[]{
			\centering
			\includegraphics[width=0.31\textwidth]{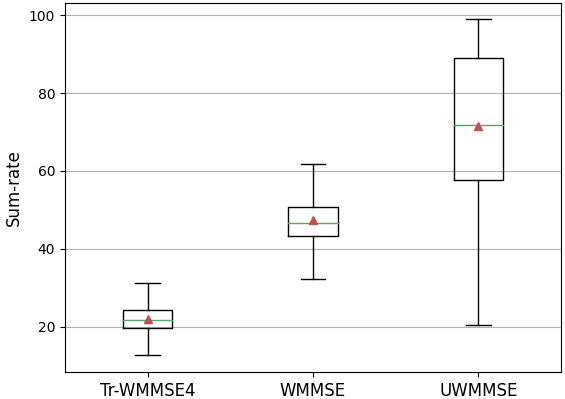}
			\label{Fig:ric}
		}	
	\subfigure[]{
			\centering
			\includegraphics[width=0.31\textwidth]{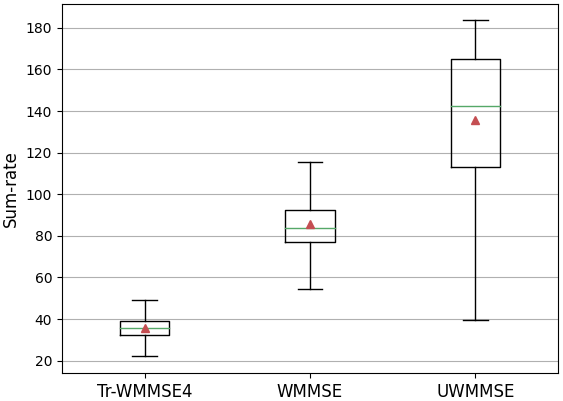}
			\label{Fig:geo}
		}
		\caption{\small Performance comparison of Truncated-WMMSE, WMMSE, and proposed UWMMSE on various channel distributions and antenna configurations. Box plots of $M=20$ user sum-rate achieved by Tr-WMMSE ($4$ iterations), WMMSE ($100$ iterations) and UWMMSE ($4$ layers) 
	    (a)~on a \emph{Rayleigh} channel with $T=5$ transmit antennas and $R=3$ receiver antennas, transmitting $d=2$ symbols;
	    (b)~on a \emph{Rician} channel with $T=3$ transmitters and $R=3$ receivers, transmitting $d=2$ symbols;
	    (c)~on a \emph{Geometric} channel with $T=3$ transmitters and $R=5$ receivers, transmitting $d=2$ symbols.
		}
		\label{Fig:comparison}
\end{figure*}

We now explain the non-linear function $\beta$ in~\eqref{E:unfold_v}.
To incorporate the power constraint in the optimization problem, the proposed UWMMSE uses a saturation non-linearity $\beta$ in its unfolded layers, ensuring that individual layer outputs obey the power constraint. 
For the multi-antenna setup that we consider here, this involves constraining the transmit beamformer $\mathbfcal{V}$ such that all its elements $\mathbf{V}_i = [\mathbfcal{V}]_{i::}$ satisfy $\Tr\left(\mathbf{V}_i^{}\mathbf{V}_i^{\top}\right) \leq P_{\mathrm{max}}$. 
To attain this, the activation $\beta$ applied to an arbitrary matrix $\mathbf{A}$ is defined as
\begin{align}\label{eq:nonlinear_beta}
    \beta(\mathbf{A}) = 
\begin{cases}
    \mathbf{A},& \text{if } \Tr\left(\mathbf{A} \mathbf{A}^{\top}\right) \leq P_{\mathrm{max}},\\
    \mathbf{A} \cdot  \frac{\sqrt{ P_{\mathrm{max}}}}{||\mathbf{A}||_F},              & \text{otherwise},
\end{cases}
\end{align}
where $||\cdot||_F$ denotes the Frobenius norm. A non-linear mapping of this form was used in the PGD based beamforming strategy of~\cite{pellaco2020deep} as the projection step. 

Having explained the inner workings of the proposed UWMMSE for a given set of parameters $\Theta = \{ \theta_a, \theta_b, \omega\}$ we now shift focus to the training of the architecture.
For fixed $\Theta$, the transmitter beamforming for CSI $\mathbfcal{H}$ is given by $\Lambda( \mathbfcal{H} ; \Theta)$ and results in a sum-rate utility of $\sum_{i=1}^M c_i(\Lambda( \mathbfcal{H} ; \Theta), \mathbfcal{H})$; [cf.~\eqref{E:optimization_problem} for $\alpha_i=1$].
Hence, we define the loss function
\begin{equation}\label{eq:loss_function}
    \ell(\Theta) = - \mathbb{E}_{\mathbfcal{H} \sim \mathcal{D}} \left[ \sum_{i=1}^M c_i(\Lambda( \mathbfcal{H} ; \Theta), \mathbfcal{H}) \right],
\end{equation}
where $\mathcal{D}$ is the channel state distribution of interest. 
Even if $\mathcal{D}$ is known, minimizing $\ell(\Theta)$ with respect to $\Theta = \{ \theta_a, \theta_b, \omega\}$ is a non-convex problem.
However, notice that $\Psi(\cdot; \theta)$ in~\eqref{E:unfold_3} and $\Phi(\cdot; \omega)$ in~\eqref{eq:H_bar} are differentiable with respect to $\theta$ and $\omega$, respectively.
Thus, given samples of $\mathbfcal{H}$ drawn from $\mathcal{D}$, we seek to minimize~\eqref{eq:loss_function} through stochastic gradient descent.
In this sense, we can claim that UWMMSE is an unsupervised method since it requires
access to samples of the channel state tensors $\mathbfcal{H}$ but does \emph{not} require access to the optimal beamformers (labels)
associated with those channels.
Further notice that a single pair of GNNs is shared by all unfolding layers, i.e., $\theta_a^{}$ and $\theta_b^{}$ do not depend on the layer index $k$. 
Such a scheme ensures that $\theta_a$ and $\theta_b$ are trained using gradient feedback that accumulates across layers and is a function of the overall optimization process. 
This arrangement creates an implicit memory through which the sequential updates of the WMMSE parameters are captured in the GNN. 
Moreover, a formulation of this form decouples training and inference architectures and thereby allows for flexibility in adding or removing unfolded layers at deployment. 
Another immediate advantage is an $\mathcal{O}(K)$ reduction in the number of trainable weights with respect to the layer-dependent alternative, making the training process less computationally expensive and less time consuming.   

\begin{remark}[Application to SISO systems]\label{R:comparison} 
Although designed for the general MIMO setting, the algorithm proposed here can be seamlessly used for power allocation in SISO wireless networks.
In such a setting, there would be no need for the neural network in~\eqref{eq:H_bar} since when $R=T=1$ the CSI is already in matrix form, thus reducing the trainable parameters only to $\Theta = \{ \theta_a, \theta_b\}$.
Similarly, intermediate variables $\mathbf{U}^{(k)}_i, \mathbf{W}^{(k)}_i, \mathbf{V}^{(k)}_i$ are reduced to scalars, but the update equations in~\eqref{E:unfold_u}-\eqref{E:unfold_v} are still valid (and computationally less demanding).
In this sense, for the SISO case, our unfolding UWMMSE highly resembles the unfolding scheme presented in~\cite{chowdhury2021unfolding}.
However, in~\cite{chowdhury2021unfolding} each layer $k$ of the model has its own independent set of GNNs $\Psi(\cdot;\theta^{(k)})$ resulting in growing complexity with increasing number of layers. 
More importantly, a model of this form requires the number of layers to be fixed at both training and inference making it inflexible at deployment.
In this respect, our more general UWMMSE framework for MIMO still presents advantages with respect to existing works even if we focus on the particular SISO setting.
\end{remark}

\section{Numerical Experiments}
\label{sec:num_exp}

In this section, we illustrate the performance of the proposed model in allocating power to multi-antenna ad-hoc wireless networks operating under various fading conditions and topologies\footnote{Code to replicate the numerical experiments presented here can be found at \href{https://github.com/ArCho48/UWMMSE_MIMO.git}{https://github.com/archo48/uwmmse\_mimo.git}.}.
A detailed description of the datasets is provided in Section~\ref{subsec:data}. 
In Section~\ref{subsec:comp}, we compare our model performance with WMMSE and its truncated version, in terms of achieved sum-rate and allocation time. 
Further, in Section~\ref{subsec:rob} we investigate robustness of our proposed model to variations in network size and channel conditions. 
Our proposed architecture is composed of $4$ unfolded WMMSE layers with two $2$-layered GCNs in each unfolded layer modeling the function $\Psi$ in~\eqref{E:unfold_3}. 
The hidden layer dimension of all GCNs is set to $5$. A maximum of $15000$ training iterations are performed with early stopping. 
Batch size is fixed at $64$ and learning rate is set to \num{1e-2}. We run $\sim 160$ test iterations with the same batch size. 
All computations were performed on an Intel Core i7-9850H CPU with 16GB RAM. 

\subsection{Datasets}\label{subsec:data}

We use randomly generated channel realizations to evaluate the model performance. 
Similar to~\cite{sun2021learning,sun2018learning,liang2019towards}, we choose the following $3$ random channel models:\\  

\noindent \textbf{Rayleigh}: For each channel matrix $\mathbf{H}_{ij}$ corresponding to the transceiver pair $ij$, we generate Rayleigh channel coefficients $[\mathbf{H}_{ij}]_{rt}$ independently for all antenna pairs $rt$ as the absolute value of real and imaginary components sampled from a standard normal distribution. 
%

\noindent \textbf{Rician}: For each channel matrix $\mathbf{H}_{ij}$ corresponding to the transceiver pair $ij$, we generate Rician channel coefficients $[\mathbf{H}_{ij}]_{rt}$ with $20$ dB $K$-factor{~\cite{tse2005fundamentals}} independently for all antenna pairs $rt$ as the absolute value of real and imaginary components sampled from a normal distribution. 

\noindent \textbf{Geometric}: A geometric channel model has a composite structure with path loss and Rayleigh fading components. 
To that end, we construct a random geometric graph in two dimensions having $M$ transceiver pairs. 
All transmitters and receivers are dropped uniformly at random at location $\mathbf{t}_i \in [-\sqrt{M}, \sqrt{M}]^2$ and $\mathbf{r}_i \in [-\sqrt{M}, \sqrt{M}]^2$. 
Path loss between transmitter $i$ and receiver $r(j)$ is then computed as a function of their corresponding physical distance $d_{ij}$. 
Incorporating Rayleigh fading, elements of the channel matrix $[\mathbf{H}_{ij}]_{rt}$ are given by
\begin{align*}
    [\mathbf{H}_{ij}]_{rt} = \frac{1}{1 + d_{ij}^2}\left|\frac{\mathscr{N}(0,1)}{\sqrt{2}} + \text{i}\frac{\mathscr{N}(0,1)}{\sqrt{2}}\right| \quad \forall r,t.
\end{align*}

\subsection{Performance comparison}\label{subsec:comp}

We compare the performance attained by our method with that of established baselines in the low-noise regime [$\sigma = \num{2.6e-5}$ in~\eqref{E:data_rate}] where the achievable capacity depends strongly on the interference between users making the power allocation process more challenging~\cite{chowdhury2021unfolding}.
We choose the following baselines for comparison:

\begin{enumerate}
    \item WMMSE \cite{shi2011iteratively} forms the baseline for our experiments. We set a maximum of $100$ iterations per sample.
    \item Truncated WMMSE (Tr-WMMSE) provides a performance lower bound to UWMMSE.  
    We fix the number of iterations to $4$ to match UWMMSE unrollings.   
\end{enumerate}
The comparisons are shown in Figure~\ref{Fig:comparison}. 
Since the channel realizations are sampled randomly, we observe significant variation in the utility values for individual samples under optimal power allocation.

The comparison shows that the performance of our method exceeds that of WMMSE, on average, for all three channel models under various antenna configurations. 
Moreover, there is a significant gap in achieved mean sum-rate between UWMMSE and Tr-WMMSE with the same number of iterations as unfolded layers in all cases, proving the effectiveness of a learning-based module in a hybrid method of this form. 
The superiority of our method can be further illustrated using the inference time comparisons. 
While WMMSE requires $\sim \mathbf{600\ msec}$ on average to converge, per sample, our method has an average inference time of $\sim \mathbf{30\ msec}$ effectively providing a $20 \times$ speedup. 
Time taken by Tr-WMMSE is $\sim \mathbf{24\ msec}$, however, the achieved performance is poor. 
Our method, therefore, exceeds the performance of WMMSE with a full set of iterations under a time complexity that is comparable to its truncated version. 

\subsection{Robustness}\label{subsec:rob}

In this section, we investigate the robustness of our proposed method to variations in channel distributions and wireless network size. 
These are realistic scenarios since the channel distribution depends largely on geographical and climatic conditions as well as the community structure (rural, urban, suburban) in a given area. 
Further, an ad-hoc wireless network may undergo modifications in size as new users join the network and existing users leave. 
It is imperative that the power allocation method be able to maintain a steady performance in spite of these variations. 
To this end, we compare the performances of two versions of our model, one of which is trained on Rayleigh and the other is trained on Rician channel realizations. 
These comparisons are performed independently on a Rayleigh and a Rician test set. Figures~\ref{Fig:ro1} and~\ref{Fig:ro2} show that both models are effective in achieving comparable performance for out-of-distribution data points. 
This observation indicates that our model can endure changes in channel conditions without significant performance degradation. 
In Figure~\ref{Fig:ro3}, we consider a slightly more challenging scenario wherein, under a Geometric channel setting, the network size varies within a fixed range. 
To evaluate the interpolation behaviour of our model, we sample only odd-number network sizes in training while the test set comprises of all integer values in the specified range. 
In this case, the Rayleigh and Rician versions of UWMMSE fail to maintain a steady performance across the range of sizes, doing slightly better for a few smaller sizes but degrading fast as the network grows. 
However, UWMMSE trained on Geometric channels is able to maintain a constant gain over WMMSE at all sizes. 
This superiority can be attributed to the GNN learning module which is regularized by the inherent graph structure of Geometric channels. 
Since real-world channel models are more aligned toward the Geometric setup composed of path loss and fading, our model has a natural advantage in dealing with more challenging scenarios that will be considered in future versions of this work.      

\section{Conclusion}

We proposed a MIMO version of UWMMSE, a hybrid learnable power allocation algorithm for wireless networks. 
The main contribution of this method is the use of low-complexity GNN-based \emph{learning modules} that leverage the connectivity structure of the transceivers to learn a mapping for faster parameter updates of an \emph{unfolded} iterative algorithm tuned to approximate a global solution of the NP-hard power allocation problem. 
It is this hybrid architecture that captures on the one hand the dynamic structure of the iterative solution while on the other hand approximates the cumbersome computational steps with learnable weights and drives the model towards faster convergence.
In the future, we will go beyond synthetic data and validate empirically the effectiveness of our proposed method on real-world datasets.

\bibliographystyle{IEEEbib}
\bibliography{strings,refs}

\end{document}